\documentclass[lettersize,journal]{IEEEtran}
\usepackage{amsmath,amsfonts}
\usepackage{algorithmic}
\usepackage{algorithm}
\usepackage{array}
\usepackage{textcomp}
\usepackage{stfloats}
\usepackage{url}
\usepackage{verbatim}
\usepackage{graphicx}
\usepackage{cite}
\usepackage{booktabs}
\usepackage{multirow}
\usepackage{rotating}
\usepackage{tabularx}
\usepackage{booktabs}
\usepackage{subcaption}
\usepackage{caption}
\usepackage{threeparttable}
\usepackage{hyperref}
\title{MSGM: A Multi-Scale Spatiotemporal Graph Mamba \\ for EEG Emotion Recognition}

\author{
    \IEEEauthorblockN{Hanwen Liu\textsuperscript{\#}}
    ,
    \IEEEauthorblockN{Yifeng Gong\textsuperscript{\#}}
    ,
    \IEEEauthorblockN{Zuwei Yan\textsuperscript{\#}}
    ,
    \IEEEauthorblockN{Zeheng Zhuang\textsuperscript{\#}}
    ,
    \IEEEauthorblockN{Jiaxuan Lu\textsuperscript{*}}
\thanks{\textsuperscript{\#} Contributed equally to this work.} 
\thanks{Hanwen Liu and Yifeng Gong are with the School of Electronics and Communication Engineering, Sun Yat-sen University, Shenzhen, 518107, China, e-mail: (liuhw56, gongyf9)@mail2.sysu.edu.cn.}
\thanks{Zuwei Yan is with the College of Communication Engineering, Jilin University, Changchun, 130012, China, e-mail: yanzw2422@mails.jlu.edu.cn.}
\thanks{Zeheng Zhuang is with the School of Pharmaceutical Sciences (Shenzhen), Sun Yat-sen University, Shenzhen, 518107, China, e-mail: zhuangzh23@mail2.sysu.edu.cn.}
\thanks{Jiaxuan Lu is with Shanghai Artificial Intelligence Laboratory, Shanghai, 200232, China, e-mail: lujiaxuan@pjlab.org.cn}
\thanks{* Corresponding Author.}
}

\begin{document}
\maketitle

\begin{abstract}
EEG-based emotion recognition struggles with capturing multi-scale spatiotemporal dynamics and ensuring computational efficiency for real-time applications. Existing methods often oversimplify temporal granularity and spatial hierarchies, limiting accuracy. To overcome these challenges, we propose the Multi-Scale Spatiotemporal Graph Mamba (MSGM), a novel framework integrating multi-window temporal segmentation, bimodal spatial graph modeling, and efficient fusion via the Mamba architecture. By segmenting EEG signals across diverse temporal scales and constructing global-local graphs with neuroanatomical priors, MSGM effectively captures fine-grained emotional fluctuations and hierarchical brain connectivity. A multi-depth Graph Convolutional Network (GCN) and token embedding fusion module, paired with Mamba's state-space modeling, enable dynamic spatiotemporal interaction at linear complexity. Notably, with just one MSST-Mamba layer, MSGM surpasses leading methods in the field on the SEED, THU-EP, and FACED datasets, outperforming baselines in subject-independent emotion classification while achieving robust accuracy and millisecond-level inference on the NVIDIA Jetson Xavier NX.
\end{abstract}


\section{Introduction}
\IEEEPARstart{E}{motion} recognition has emerged as a critical research frontier with far-reaching implications for human-computer interaction, mental health monitoring, and neuroscientific exploration\cite{9153955}\cite{cheng2024eeg}\cite{guo2022transformer}. The ability to decode emotional states in real-time promises to revolutionize intelligent systems by enhancing user adaptability and bolstering clinical applications through early detection and management of emotional disorders\cite{jafari2023emotion}\cite{xiao2025multi}. As these capabilities become increasingly vital in healthcare and artificial intelligence, there is an urgent need for robust, efficient, and neurophysiologically grounded approaches to overcome both theoretical complexities and practical deployment challenges\cite{vafaei2025transformers}.

Electroencephalography (EEG) stands out as a premier modality for emotion recognition, owing to its unparalleled capacity to non-invasively record brain activity with high temporal resolution, directly capturing the neural signatures of emotional processes\cite{yao2024emotion}. Unlike indirect proxies such as facial expression analysis or speech intonation, EEG offers immediate access to the brain’s dynamic responses, making it ideally suited for applications requiring precision and responsiveness. However, the efficacy of EEG-based emotion recognition is tempered by significant hurdles: the signals’ susceptibility to noise, their spatial heterogeneity across brain regions\cite{liu2024comprehensive}, and their complex temporal dynamics, which span short-term fluctuations and long-term trends yet are often inadequately modeled by fixed-scale approaches\cite{liu2021review}.

The progression of EEG-based emotion recognition encapsulates a diverse methodological landscape\cite{xiao2024multisource}. Traditional techniques leaned heavily on manually engineered features, such as wavelet transforms \cite{gosala2023wavelet}\cite{jmail2018integration}, augmented by neuroscientific priors to distill interpretable patterns. While these methods afford a degree of transparency, their reliance on labor-intensive processes and specialized expertise limits scalability, tethering them to a conventional paradigm. In contrast, modern strategies harness deep learning architectures, including convolutional neural networks (CNNs)\cite{Song2020EEG}, recurrent neural networks (RNNs)\cite{7112127}, and Transformers\cite{vafaei2025transformers}, to automate feature extraction and elevate performance. Within this advanced framework, features like power spectral density (PSD), relative power spectral density (rPSD), and differential entropy (DE) are commonly integrated\cite{yan2024bridge}, marrying neuroscientific insight with computational prowess. Yet, despite these advances, significant limitations persist in fully capturing the complexity of emotional dynamics, raising critical questions about the adequacy of current approaches.

Despite these advancements, contemporary EEG-based emotion recognition approaches remain hampered by critical deficiencies that our research addresses with the Multi-scale Spatiotemporal Graph Mamba (MSGM). To bridge these gaps, our MSGM framework integrates a novel graph-based Mamba structure with multi-scale spatiotemporal analysis to comprehensively model the intricate dynamics of EEG signals. Firstly, emotional states manifest temporal dynamics across multiple scales, short-term fluctuations reflecting immediate responses and long-term trends signifying sustained shifts, yet prevailing models rely on fixed temporal windows \cite{ding2023masa}, neglecting this multi-granular complexity; MSGM tackles this by unifying multi-scale temporal analysis through a multi-window sliding strategy, extracting rPSD features from seven frequency bands to capture both fleeting shifts and prolonged trends via its Temporal Multi-scale Feature Extraction module. 

Secondly, the spatial topology of emotional processing in the brain is inherently distributed and hierarchical\cite{li2019regional}, encompassing both global functional connectivity and localized regional interactions\cite{9979692}\cite{10025569}, a sophistication that single-scale spatial representations fail to encapsulate\cite{9892411}; MSGM addresses this with hierarchical spatial graphs, using neuroanatomical priors to build adaptive global graphs and local subgraphs, fused via multi-depth GCNs and token embeddings through its Spatial Multi-scale Prior Information Initialization and Spatiotemporal Feature Adaptive Fusion modules. 

Thirdly, the pursuit of high recognition accuracy often incurs substantial computational costs \cite{ye2022twostage}, hindering real-time deployment on resource-constrained edge devices essential for clinical and consumer applications \cite{8890664}. Recently, Mamba has garnered attention for its linear-time sequence modeling via selective state spaces, providing superior efficiency and scalability over Transformers, especially for long sequences \cite{gu2023mamba}. In EEG-based emotion recognition, initial explorations have shown the efficacy of Mamba's state-space models (SSMs) in handling complex neural dynamics with lower overhead; for example, MS-iMamba uses an inverted structure for spatiotemporal fusion to achieve high benchmark accuracies \cite{zhou2024multi}, and Global Context MambaVision combines SSMs with contextual modeling to improve efficiency and performance \cite{10890602}. Drawing on these foundations, MSGM addresses these challenges through a graph-based Mamba structure, enabling edge efficiency with inference times below 151 ms on the NVIDIA Jetson Xavier NX.

Our MSGM framework advances EEG-based emotion recognition with the following contributions:

1) We propose the MSGM network to address subject-independent emotion classification, decoding complex EEG emotional dynamics with high precision.

2) We introduce the Temporal Multi-scale Feature Extraction, Spatial Multi-scale Prior Information Initialization, and Spatiotemporal Feature Adaptive Fusion modules to enhance modeling of temporal granularity and spatial connectivity.

3) MSGM delivers superior performance on the SEED\cite{zheng2015investigating}, THU-EP\cite{hu2022similar}, and FACED datasets\cite{chen2023large}, surpassing baselines such as DGCNN\cite{8320798} in subject-independent settings. Notably, with only a single MSST-Mamba layer, it outperforms leading methods in the field on the same datasets.

4) Deployed on the NVIDIA Jetson Xavier NX, MSGM delivers real-time inference within 151 ms, enabling efficient performance on resource-constrained edge devices.

\section{Related Works}

\subsection{Graph Neural Networks in EEG Analysis}

Graph neural networks (GNNs) have become a cornerstone for modeling the spatial topology of electroencephalography (EEG) signals, leveraging their inherent graph structure where channels represent nodes and inter-channel relationships form edges. Spectral GNNs, such as ChebyNet\cite{xinchebynet}, employ Chebyshev polynomials to approximate graph Laplacian filters, enabling efficient spatial feature extraction across EEG channels. Similarly, graph convolutional networks (GCNs) with first-order approximations, as in DGCNN, dynamically learn adjacency matrices to capture spatial dependencies, achieving robust emotion recognition performance, demonstrating the feasibility of this direction. RGNN \cite{RGNN} further incorporates neuroscientifically inspired constraints into its adjacency matrix, enhancing biological interpretability and achieving an impressive accuracy on SEED, demonstrating remarkable progress in the field.

Likewise, BiDANN\cite{BiDANN} excellently leverages hemispheric asymmetry for emotion recognition, proving the potential of GNN-based approaches. These methods excel in modeling local spatial patterns, reflecting functional connectivity among brain regions. However, their reliance on static or single-scale adjacency matrices often overlooks the dynamic, hierarchical interactions across the brain’s distributed emotional networks\cite{lu2024comprehensive}. Moreover, many GNN-based approaches, such as GCB-Net\cite{zhang2019gcb}, utilize averaged features as node attributes, sacrificing critical temporal context essential for capturing emotional dynamics. To address this limitation, adaptive spatial modeling that evolves during training and incorporates multi-scale dependencies becomes essential, a challenge our MSGM effectively resolves through its bimodal spatial graph structure, initialized with neuroanatomical priors and dynamically refined.

\subsection{Multi-Scale Representation Learning}

Multi-scale representation learning has emerged as a potent strategy for time series analysis, adept at capturing both fine-grained details and overarching trends, with growing applications in EEG signal processing. In domains like time series prediction, methods such as TimesNet\cite{wu2023timesnet} transform one-dimensional sequences into two-dimensional tensors across multiple scales, modeling intra- and inter-periodic variations with convolutional kernels. In EEG emotion recognition, AMCNN-DGCN\cite{wang2020linking} employs multi-scale convolutional blocks to extract temporal features, circumventing manual feature engineering, while Pathformer\cite{chen2024pathformer} integrates dual attention mechanisms across varying time resolutions to balance local and global dependencies. These approaches demonstrate the power of multi-scale frameworks in addressing the limitations of fixed-scale models, which often fail to represent the diverse temporal dynamics of emotional states.

Notably, Visual-to-EEG Cross-Modal Knowledge Distillation (CKD)\cite{zhang2022visual} effectively leverages TCN to extract temporal features from EEG, achieving an impressive RMSE of 0.064 on MAHNOB-HCI, proving the viability of multi-scale-inspired methods. Similarly, DMATN\cite{Wang2021} excels in cross-subject EEG analysis with multi-source feature extraction, highlighting the promise of this direction. Nevertheless, their application in EEG remains nascent, with most methods either focusing on a single temporal granularity or neglecting spatial integration\cite{5444455}. This shortfall restricts their ability to holistically model the brain’s multi-scale neural oscillations underlying emotions. Our MSGM advances this paradigm by introducing a multi-window sliding segmentation strategy, capturing both short-term continuity and long-term evolutionary patterns, thereby enhancing temporal granularity within a biologically informed framework.

\subsection{Spatio-Temporal Fusion in EEG Emotion Recognition}

The fusion of spatial and temporal features has gained traction in EEG emotion recognition, aiming to harness the brain’s complex spatiotemporal dynamics. Hybrid architectures, such as Conformer\cite{song2022eeg}, combine convolutional neural networks (CNNs) with transformers to integrate short-term spatial patterns and long-term temporal dependencies, achieving promising classification results. Similarly, ASTDF-Net\cite{gong2023astdf} employs dual-stream attention to learn a joint spatiotemporal subspace, while SGCN-LSTM\cite{feng2022eeg} hybrids pair graph convolutions with recurrent units to model spatial topology and temporal continuity. Notably, Soleymani’s work on continuous emotion detection\cite{7112127} excels by using LSTM-RNN to fuse EEG and facial expression features, achieving a robust Pearson correlation of 0.48 on MAHNOB-HCI\cite{soleymani2011multimodal}, proving the efficacy of spatiotemporal integration.

Likewise, BiDANN brilliantly combines LSTM with adversarial training to capture hemispheric dynamics, yielding an impressive accuracy on SEED, showcasing the potential of this approach. DMATN further demonstrates excellence by integrating multi-source EEG features, reinforcing the viability of this direction. These methods offer significant advances over isolated spatial or temporal approaches, yet they frequently process these dimensions in parallel branches, limiting interactive feature integration. Moreover, their computational complexity, often driven by attention mechanisms or deep convolutional stacks, renders them impractical for real-time applications on resource-constrained edge devices, which is a critical requirement for clinical and consumer use. Additionally, few models incorporate biologically inspired mechanisms to reflect the brain’s hierarchical emotional processing or prioritize efficiency alongside accuracy. Our MSGM addresses these deficiencies with a unified spatiotemporal fusion module that blends multi-depth GCNs with token embeddings for dynamic interaction, while leveraging the efficient Mamba architecture to capture global and local dependencies, ensuring both biological interpretability and practical applicability.

\begin{figure*}[h!]
\centering
\includegraphics[width=\textwidth]{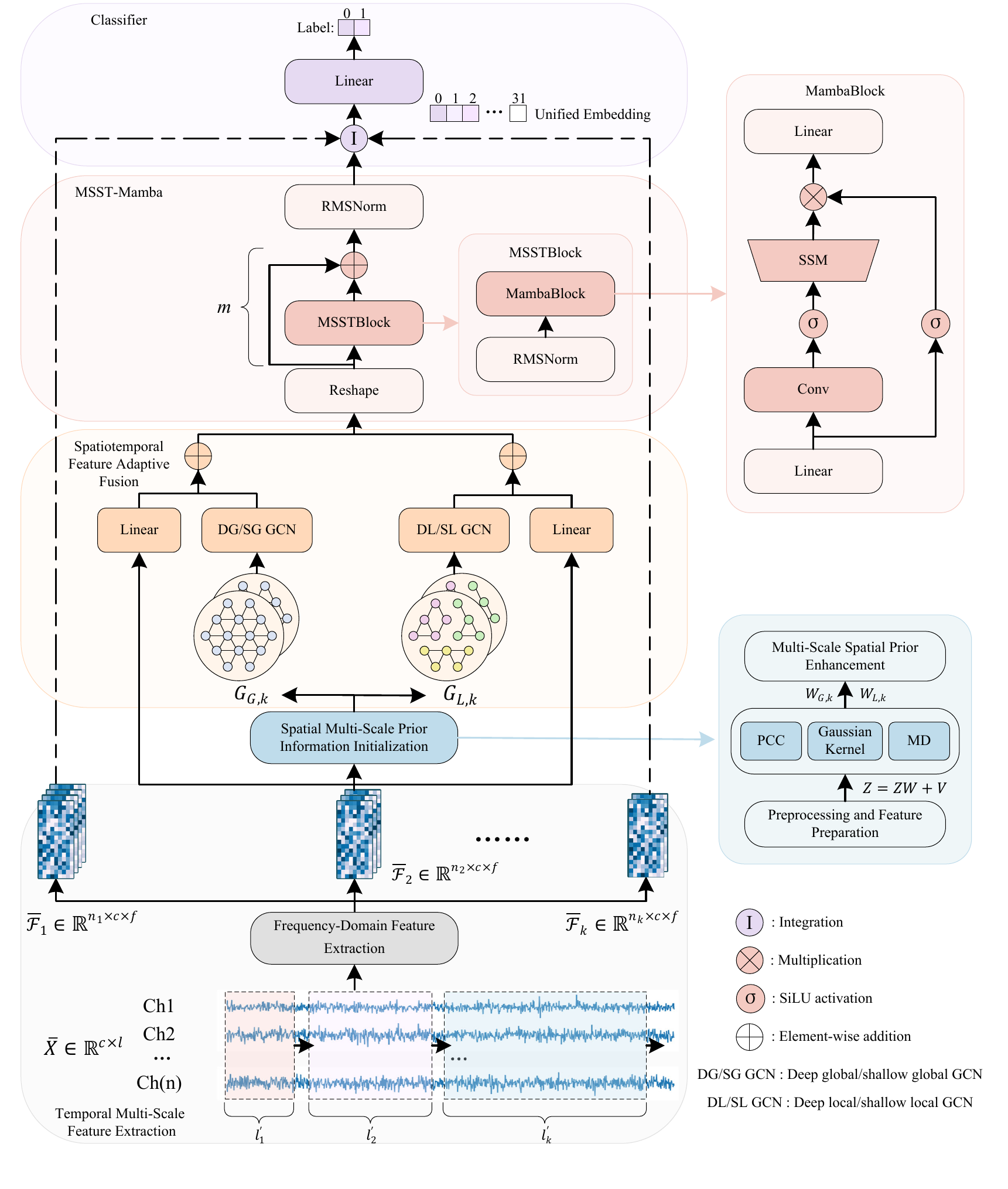}
\caption{The framework of MSGM. The multi-scale feature tensors from Temporal Multi-Scale Feature Extraction module are used as the input to Spatial Multi-Scale Prior
Information Initialization module that will transfer tensors into global graphs and local graphs. Then Spatiotemporal Feature Adaptive Fusion module extract dynamic spatial relationships among EEG channels via GCNs and temporal fusion. The MSST-Mamba block then processes the input tensors, followed by the final Classifier module.}
\label{fig:the_framework_of_MSGM}
\end{figure*}

\section{Method}
In this section, we present the details of the proposed method, which comprises temporal multi-scale feature extraction, spatial multi-scale prior information initialization, spatiotemporal feature adaptive fusion, MSST-Mamba and classifier. The overall architecture of the proposed method is depicted in Figure~\ref{fig:the_framework_of_MSGM}.

\subsection{Temporal Multi-Scale Feature Extraction}
To effectively analyze emotional states from EEG signals, a multi-scale feature extraction process is employed. This section details the three key stages: multi-scale temporal segmentation, frequency-domain feature extraction using relative power spectral density (rPSD), and multi-scale feature tensor generation.

\subsubsection{Multi-Scale Temporal Segmentation}
The raw EEG signal, denoted as \( X \in \mathbb{R}^{c \times L} \) where \( c \) represents the number of channels and \( L \) indicates the total number of time samples, is processed through a two-level segmentation method to effectively capture its multi-scale temporal dynamics. In the initial stage, known as first-level segmentation, the signal \( X \) is segmented into larger portions \( \overline{X} \in \mathbb{R}^{c \times l} \) by applying a sliding window of length \( l = 20sec = 20*f_s \) with a hop step \( s = 4sec = 4 * f_s \), where \( f_s \) represents the sampling frequency of EEG signals, resulting in overlapping segments that encompass wider temporal contexts within the EEG data. Following this, the second-level segmentation takes each of these larger segments \( \overline{X} \) and further divides them into smaller sub-segments using \( k \) distinct sliding windows, each characterized by specific lengths \( l'_k \) and hop steps \( s'_k \) for \( k = 1, 2, 3, \ldots, k \). This process yields \( k \) sets of sub-segments \( \tilde{X}_k \in \mathbb{R}^{c \times l'_k} \), with each set offering a unique temporal resolution of the brain activity contained within the same larger segment.

\subsubsection{Frequency-Domain Feature Extraction}
For each sub-segment \( \tilde{X}_k \) derived from the \( k \) different time window lengths, spectral features are extracted by applying the Fast Fourier Transform (FFT) to each channel. The signal is decomposed into seven frequency bands: delta (1-4 Hz), theta (4-8 Hz), alpha (8-12 Hz), low beta (12-16 Hz), beta (16-20 Hz), high beta (20-28 Hz), and gamma (30-45 Hz). The relative power spectral density (rPSD) is then computed for each band using Welch’s method, yielding a feature matrix \( F_k \in \mathbb{R}^{c \times f} \) for each sub-segment \( \tilde{X}_k \), where f = 7. Experimental comparisons with alternatives like PSD and differential entropy (DE) confirmed that rPSD provides superior performance in distinguishing emotional states. These rPSD values are later used as node attributes in the graph representation.

\subsubsection{Multi-Scale Feature Tensor Generation}
The rPSD features extracted from the previous step are organized into \( k \) distinct feature tensors, each corresponding to one of the temporal scales defined by the window sizes \( l'_1, l'_2, \ldots, l'_k \). For each scale \( k \), the resulting feature tensor is structured as \( \mathcal{F}_k \in \mathbb{R}^{b \times n_k \times c \times f} \), where \( b \) is the batch size, \( n_k \) is the number of segments for the \( k \)-th window size. This multi-scale tensor representation preserves the temporal information at different granularities and provides a comprehensive spatio-temporal characterization of the EEG signals.

\subsection{Spatial Multi-Scale Prior Information Initialization}
This subsection outlines a method for initializing spatial prior information across multiple scales in EEG analysis. The approach involves three key steps: preprocessing and feature preparation to extract relevant EEG features, construction of global and local graphs to model channel interactions, and enhancement of multi-scale spatial priors to improve the representation of connectivity patterns.

\subsubsection{Preprocessing and Feature Preparation}
Using the preprocessed multi-scale feature tensor \( \mathcal{F}_k \in \mathbb{R}^{b \times n_k \times c \times f} \), spatial graphs are constructed to represent channel interactions. To establish a consistent graph structure across the batch, we compute the average of \(\mathcal{F}_k\) over the batch dimension, yielding \(\overline{\mathcal{F}}_k \in \mathbb{R}^{n_k \times c \times f}\). This averaging reduces computational complexity while preserving common spatial patterns within the data. Subsequently, \(\overline{\mathcal{F}}_k\) is reshaped into a matrix \(Z \in \mathbb{R}^{c \times (f n_k)}\) by flattening the sequence and feature dimensions. To adaptively combine features across frequency bands and time segments, a learnable transformation is applied:

\begin{align}
Z = Z W + V, \tag{1} \label{eq:1}
\end{align}

where \( W \in \mathbb{R}^{(f n_k) \times n_k} \) is a trainable weight matrix initialized using Xavier uniform initialization to ensure stable gradient flow during training, and \( V \in \mathbb{R}^{c \times n_k} \) is a bias matrix initialized as zeros. This transformation enables the model to learn optimal feature combinations, enhancing its sensitivity to emotional patterns embedded in the EEG signals.

\subsubsection{Construction of Global and Local Graphs}
\begin{figure}[t!]
\centering
\includegraphics[width=\columnwidth]{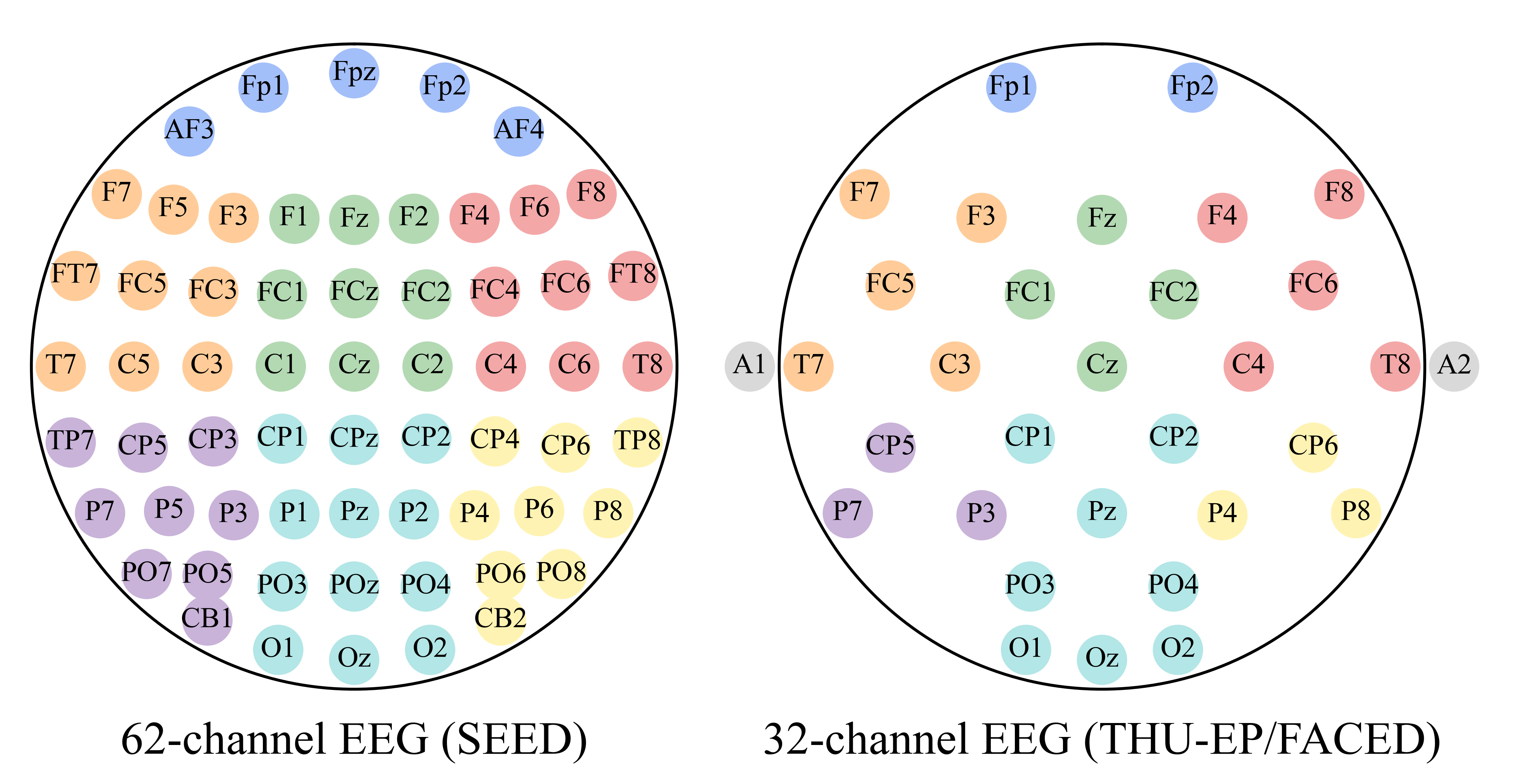}
\caption{The division method of 62-channel and 32-channel EEG. The same color represents the same region.}
\label{fig:62-channel_and_32-channel_EEG}
\end{figure}

At each scale \( k \), two graphs are defined: a global graph \(G_{G,k} = (U, E_{G,k})\) and a local graph \(G_{L,k} = (U, E_{L,k})\). Both graphs share the same node set \(U = \{u_1, u_2, \dots, u_c\}\), where each node \(u_i\) corresponds to an EEG channel, and the feature vector for node \(u_i\), denoted \(\mathbf{u}_{i,k} \in \mathbb{R}^{n_k}\), is extracted directly from \(Z\). The global adjacency matrix \(W_{G,k} \in \mathbb{R}^{c \times c}\) is constructed using a hybrid metric that integrates the Pearson Correlation Coefficient (PCC) and Manhattan Distance (MD) to eliminate weak or noisy connections while retaining meaningful spatial relationships. The PCC, \(\kappa_{i,j,k}\), is calculated after normalizing the feature vectors by subtracting their mean and dividing by their standard deviation, with a small constant (\(1e-6\)) added to the denominator to avoid division by zero in cases of constant features. The MD is computed as \(d_{i,j,k} = \|\mathbf{u}_{i,k} - \mathbf{u}_{j,k}\|_1\), capturing the absolute differences between feature vectors. The weights in \(W_{G,k}\) are then defined as:

\begin{align}
w_{ij}^{G,k} &=
\begin{cases}
\exp\left(-\dfrac{\|\mathbf{u}_{i,k} - \mathbf{u}_{j,k}\|_2^2}{2\sigma^2}\right), & \begin{aligned}
&\text{if } \kappa_{i,j,k} \geq \kappa_\theta \\
&\text{and } d_{i,j,k} \leq d_\theta,
\end{aligned} \\
0, & \text{otherwise},
\end{cases} \tag{2} \label{eq:2}
\end{align}

where \(\sigma\) is the Gaussian kernel bandwidth, adaptively set to \((\mu_d + \sigma_d)/2\)—the average of the mean (\(\mu_d\)) and standard deviation (\(\sigma_d\)) of Euclidean distances across all node pairs—unless specified otherwise. The thresholds \(\kappa_\theta\) and \(d_\theta\) are set to the 75th percentile of PCC values and the 25th percentile of MD values, respectively, ensuring data-driven robustness without requiring manual tuning. In contrast, the local adjacency matrix \(W_{L,k} \in \mathbb{R}^{c \times c}\) restricts connectivity to channels within predefined scalp regions (see Figure~\ref{fig:62-channel_and_32-channel_EEG}), defined as:

\begin{align}
w_{ij}^{L,k} =
\begin{cases}
w_{ij}^{G,k}, & \text{if } u_i \text{ and } u_j \text{ are in the same region}, \\
0, & \text{otherwise}.
\end{cases} \tag{3} \label{eq:3}
\end{align}

This dual-graph strategy effectively encapsulates both extensive inter-channel dependencies and localized interactions, forming a comprehensive spatial prior for EEG analysis.

\subsubsection{Multi-Scale Spatial Prior Enhancement}
To enhance the multi-scale spatial priors, the global and local adjacency matrices \(W_{G,k}\) and \(W_{L,k}\) at each scale \(k\) are duplicated and stacked along a new dimension, resulting in tensors \(G_{G,k} = \begin{bmatrix} W_{G,k} \\ W_{G,k} \end{bmatrix}\) and \(G_{L,k} = \begin{bmatrix} W_{L,k} \\ W_{L,k} \end{bmatrix}\), both of shape \((2, c, c)\). Although the duplicated graphs are identical in this initial setup, this structure provides flexibility for subsequent layers to apply distinct transformations or attention mechanisms, potentially enriching the representation of spatial relationships.

\subsection{Spatiotemporal Feature Adaptive Fusion}
This subsection introduces the spatiotemporal feature adaptive fusion module, which captures dynamic spatial relationships among EEG channels for emotion analysis by integrating multi-depth Graph Convolutional Networks (GCNs) and temporal fusion via token embeddings.

\subsubsection{Adaptive Graph Encoding with Multi-Depth GCNs}
The core of the spatiotemporal feature adaptive fusion module leverages four distinct Graph Encoders, each implemented using ChebyNet, a variant of GCN that employs Chebyshev polynomials to approximate spectral graph convolutions. The ChebyNet formulation is expressed as:

\begin{align}
\Phi_g(F, A) = \sigma \left( \sum_{i=0}^{I-1} \theta^i T_i(\hat{L}) F - B \right), \tag{4} \label{eq:4}
\end{align}

where \( F \in \mathbb{R}^{(b \cdot n_k) \times c \times f} \) is the input feature matrix, \( A \in \mathbb{R}^{c \times c} \) is the adjacency matrix, \( \sigma \) denotes the ReLU activation function, \( \theta^i \) are learnable parameters, \( T_i \) represent Chebyshev polynomials, \( \hat{L} = -D^{-\frac{1}{2}} A D^{-\frac{1}{2}} \) is the normalized Laplacian with degree matrix \( D \), \( B \) is the bias term, and \( I \) is the polynomial order controlling the receptive field. This formulation allows the GCNs to capture spatial dependencies efficiently by approximating the graph's spectral properties.

These four Graph Encoders process the graphs in the following manner. The Shallow Global GCN (\(\Phi_{g,\text{shallow}}^G\)) applies a shallow GCN to the first global graph \( G_{G,k}^{(1)} \). In contrast, the Deep Global GCN (\(\Phi_{g,\text{deep}}^G\)) employs a deeper GCN on the second global graph \( G_{G,k}^{(2)} \). Similarly, the Shallow Local GCN (\(\Phi_{g,\text{shallow}}^L\)) processes the first local graph \( G_{L,k}^{(1)} \) with a shallow GCN, while the Deep Local GCN (\(\Phi_{g,\text{deep}}^L\)) applies a deeper GCN to the second local graph \( G_{L,k}^{(2)} \). 

\subsubsection{Spatiotemporal Fusion Via Token Embeddings}
A linear projection layer, \(\text{LP}(\cdot)\), transforms the flattened input features \(\Gamma(\mathcal{F}_k) \in \mathbb{R}^{(b \cdot n_k) \times (c \cdot f)}\) into a base embedding \( H_{g-\text{base},k} \in \mathbb{R}^{(b \cdot n_k) \times h} \), where \( h \) is the hidden dimension, providing a non-filtered representation of the input graph.

The outputs from the GCNs and the base embedding are combined separately for the global and local graphs at each scale \( k \). The Global Graph Embedding is computed as:

\begin{align}
\begin{aligned}
s_{G,k} = \text{mean}\big( H_{g-\text{base},k}, \Phi_{g,\text{shallow}}^G(\mathcal{F}_k, A_{G,k}^{(1)}), \\
\Phi_{g,\text{deep}}^G(\mathcal{F}_k, A_{G,k}^{(2)}) \big),
\end{aligned} \tag{5} \label{eq:5}
\end{align}

where \( s_{G,k} \in \mathbb{R}^{(b \cdot n_k) \times h} \). Similarly, the Local Graph Embedding is defined as:

\begin{align}
\begin{aligned}
s_{L,k} = \text{mean}\big( H_{g-\text{base},k}, \Phi_{g,\text{shallow}}^L(\mathcal{F}_k, A_{L,k}^{(1)}), \\
\Phi_{g,\text{deep}}^L(\mathcal{F}_k, A_{L,k}^{(2)}) \big),
\end{aligned} \tag{6} \label{eq:6}
\end{align}

where \( s_{L,k} \in \mathbb{R}^{(b \cdot n_k) \times h} \). These embeddings are generated for each temporal scale, producing a set of global and local token embeddings \(\{ s_{G,k}, s_{L,k} \}\) that encapsulate multi-view spatial representations. These tokens are subsequently passed to the MSST-Mamba, which learns temporal dependencies across scales, effectively integrating both spatial and temporal patterns present in the EEG data.

\subsection{MSST-Mamba}
The MSST-Mamba module constitutes a crucial component within a broader framework designed to capture the multi-scale spatiotemporal dynamics of electroencephalogram (EEG) signals for emotion recognition. The MSST-Mamba module processes an input tensor through a stack of \( m \) MSSTBlock layers, followed by a final normalization step. Given an input \( x_{\text{in}} \in \mathbb{R}^{b \times n_k \times h} \) reshaped from the preceding spatiotemporal feature adaptive fusion module, the module’s computation can be expressed as:
\begin{align}
x_m = \text{MSSTBlock}_m(\text{RMSNorm}(x_{m-1})) + x_{m-1}, \tag{7} \label{eq:7}
\end{align}
\begin{align}
x_{\text{out}} = \text{RMSNorm}(x_m), \tag{8} \label{eq:8}
\end{align}
where \( m = 1, 2, \ldots, M \) denotes the number of layers in MSST-Mamba module, and \( h \) is the embedding dimension. Each \( \text{MSSTBlock}_m \) encapsulates a MambaBlock with a residual connection, while RMSNorm ensures numerical stability across layers. The input tensor \( x_{\text{in}} \) is sequentially transformed through the stack, culminating in a stabilized output \( x_{\text{out}} \) of the same shape.

\subsubsection{MSSTBlock}
The MSSTBlock, the foundational unit of MSST-Mamba, integrates normalization, temporal modeling, and a residual connection:
\begin{align}
z_{\text{res}} = \text{MambaBlock}(\text{RMSNorm}(z)) + z, \tag{9} \label{eq:9}
\end{align}
where \( \text{RMSNorm} \) and \( \text{MambaBlock} \) preserve the shape \( \mathbb{R}^{b \times n_k \times h} \). The residual connection ensures robust gradient flow and feature retention across layers.

\subsubsection{RMSNorm}
Root Mean Square Normalization (RMSNorm) stabilizes training by standardizing the input tensor. For an input \( z \in \mathbb{R}^{b \times n_k \times h} \), RMSNorm is defined as:
\begin{align}
z_{\text{norm}} = z \cdot \left( \sqrt{\frac{1}{h} \sum_{i=1}^{h} z_{:, :, i}^2 + \epsilon} \right)^{-1} \cdot w, \tag{10} \label{eq:10}
\end{align}
where \( w \in \mathbb{R}^{h} \) is a learnable weight vector initialized to ones, and \( \epsilon = 10^{-5} \) prevents division by zero. This normalization mitigates scale variability, enhancing learning in deep architectures.

\subsubsection{MambaBlock}
The MambaBlock performs temporal modeling within the MSSTBlock. Given a normalized input \( z_{\text{norm}} \in \mathbb{R}^{b \times n_k \times h} \), it first applies a linear projection to an expanded dimension:
\begin{align}
z_{\text{proj}} = \text{Linear}(z_{\text{norm}}) \in \mathbb{R}^{b \times n_k \times 2 d_{\text{inner}}}, \tag{11} \label{eq:11}
\end{align}
where \( d_{\text{inner}} = \text{expand} \times h \). The projected tensor is split into two components, \( u \) and \( \text{res} \), each of shape \( \mathbb{R}^{b \times n_k \times d_{\text{inner}}} \).

The \( u \) component undergoes a depthwise 1D convolution to capture local temporal dependencies:
\begin{align}
u_{\text{conv}} = \text{Conv1D}(\text{Reshape}(u))[:, :, :n_k] \in \mathbb{R}^{b \times d_{\text{inner}} \times n_k}, \tag{12} \label{eq:12}
\end{align}
where the reshape transforms \( u \) from \( (b, n_k, d_{\text{inner}}) \) to \( (b, d_{\text{inner}}, n_k) \). The convolution uses a kernel size of 4 with padding 3, trimmed to preserve the sequence length \( n_k \). The output is reshaped to \( (b, n_k, d_{\text{inner}}) \) and activated using SiLU:
\begin{align}
u_{\text{act}} = \text{SiLU}(u_{\text{conv}}). \tag{13} \label{eq:13}
\end{align}

The activated tensor \( u_{\text{act}} \) is then processed by the selective state-space model (SSM), yielding \( u_{\text{ssm}} = \text{SSM}(u_{\text{act}}) \in \mathbb{R}^{b \times n_k \times d_{\text{inner}}} \). Meanwhile, the \( \text{res} \) component is activated separately:
\begin{align}
\text{res}_{\text{act}} = \text{SiLU}(\text{res}). \tag{14} \label{eq:14}
\end{align}
The SSM output is modulated by the activated residual:
\begin{align}
u_{\text{mod}} = u_{\text{ssm}} \cdot \text{res}_{\text{act}}. \tag{15} \label{eq:15}
\end{align}
Finally, a linear projection maps the result back to the embedding dimension:
\begin{align}
z_{\text{mamba}} = \text{Linear}(u_{\text{mod}}) \in \mathbb{R}^{b \times n_k \times h}. \tag{16} \label{eq:16}
\end{align}

\subsubsection{Selective State-Space Model (SSM)}
The SSM within the MambaBlock efficiently models long-range dependencies. Starting with \( u_{\text{act}} \in \mathbb{R}^{b \times n_k \times d_{\text{inner}}} \), it projects to SSM parameters:
\begin{align}
u_{\text{dbl}} = \text{Linear}(u_{\text{act}}) \in \mathbb{R}^{b \times n_k \times (dt_{\text{rank}} + 2 d_{\text{state}})}, \tag{17} \label{eq:17}
\end{align}
which splits into \( \delta \in \mathbb{R}^{b \times n_k \times dt_{\text{rank}}} \), \( B \), and \( C \), both in \( \mathbb{R}^{b \times n_k \times d_{\text{state}}} \). The time-step parameter is computed as:
\begin{align}
\Delta = \text{Softplus}(\text{Linear}(\delta)) \in \mathbb{R}^{b \times n_k \times d_{\text{inner}}}, \tag{18} \label{eq:18}
\end{align}
where \( \text{Softplus}(x) = \ln(1 + e^x) \) ensures positivity. A learnable matrix \( A \in \mathbb{R}^{d_{\text{inner}} \times d_{\text{state}}} \), initialized as \( A = -\exp(\text{A\_log}) \), and a diagonal \( D \in \mathbb{R}^{d_{\text{inner}}} \), initialized to ones, are used.

The selective scan updates the hidden state \( x_t \) and computes the output \( y_t \) over \( t = 1, 2, \ldots, n_k \):
\begin{align}
x_t = \exp(\Delta_t \cdot A) \cdot x_{t-1} + \Delta_t \cdot B_t \cdot v_t, \tag{19} \label{eq:19}
\end{align}
\begin{align}
y_t = C_t \cdot x_t + D \cdot v_t, \tag{20} \label{eq:20}
\end{align}
where \( v_t = u_{\text{act}}[:, t, :] \), \( \Delta_t \), \( B_t \), and \( C_t \) are slices at time \( t \), and \( x_0 = 0 \in \mathbb{R}^{b \times d_{\text{inner}} \times d_{\text{state}}} \). The final output \( y \in \mathbb{R}^{b \times n_k \times d_{\text{inner}}} \) stacks \( y_t \) and adds a skip connection with \( D \), capturing both short- and long-term dependencies with linear complexity in \( n_k \).

\subsection{Classifier}
After processing through the MSST-Mamba module, the global and local \(x_{\text{out}}\) are mean-pooled along the sequence dimension and L2-normalized, then averaged to produce scale-specific representations, which are fused across all \(k\) scales via mean pooling to form a unified embedding \(x_{\text{final}} \in \mathbb{R}^{b \times h}\). This embedding captures multi-scale spatiotemporal information from the EEG signals.

The final classification output, \(\hat{y} \in \mathbb{R}^{b \times d_{\text{out}}}\), where \(d_{\text{out}}\) represents the number of emotion classes, is generated by a linear classifier applied to the unified embedding:

\begin{align}
\hat{y} = x_{\text{final}} W + b \tag{21} \label{eq:21}
\end{align}

Here, \(W \in \mathbb{R}^{h \times d_{\text{out}}}\) and \(b \in \mathbb{R}^{d_{\text{out}}}\) are the learnable weights and bias, respectively. This linear layer maps the multi-scale embedding to the logit space, producing logits that can be transformed into a probability distribution over emotion classes using the softmax function during inference or training with a cross-entropy loss.

\section{Experiment And Results}
\subsection{Datasets}
To assess the performance of our proposed model, we conducted comprehensive experiments utilizing five publicly available datasets: the SJTU Emotion EEG Dataset (SEED)\cite{zheng2015investigating}, the Emotion Profiles dataset (THU-EP)\cite{hu2022similar}, and its expanded counterpart, the FACED dataset\cite{chen2023large}.

The SEED dataset, developed by Shanghai Jiao Tong University’s BCMI laboratory, contains EEG recordings from 15 native Chinese participants (7 males, 8 females; mean age: 23.27 years). These subjects watched 15 Chinese film clips, each lasting about 4 minutes, selected to evoke three emotional states: positive, neutral, and negative (five clips per category). Following each clip, participants rated their emotions based on valence and arousal. Brain activity was captured using a 62-channel electrode cap configured per the 10-20 system, with signals recorded at a 1000 Hz sampling rate. The data was preprocessed with a 0.3–50 Hz bandpass filter to enhance signal quality for emotion analysis.

The THU-EP dataset includes EEG data from 80 college students (50 females, 30 males; aged 17–24, mean: 20.16 years) exposed to 28 video clips averaging 67 seconds each. These clips were designed to trigger nine emotions: anger, disgust, fear, sadness, amusement, joy, inspiration, tenderness, and neutral, with four clips for neutral and three for each of the others. The experiment was divided into seven blocks of four trials, with participants solving 20 arithmetic problems between blocks to reset their emotional baseline. After each clip, subjects self-reported scores for arousal, valence, familiarity, and liking. EEG signals were recorded using the NeuSen.W32 wireless system with 32 channels at a 250 Hz sampling rate, preprocessed with a 0.05–47 Hz bandpass filter, and cleaned via independent component analysis (ICA) to remove artifacts.

The FACED dataset builds on THU-EP, expanding to 123 subjects by adding 43 participants to the original 80, while retaining the same experimental framework. It employs the identical 28 video clips to elicit the nine emotions from THU-EP, following the same seven-block, four-trial structure with arithmetic tasks between blocks. Post-clip self-reports of emotional scores mirror THU-EP’s methodology. EEG data was collected with the 32-channel NeuSen.W32 system at 250 Hz, and preprocessing aligns with THU-EP, using a 0.05–47 Hz bandpass filter and ICA for artifact removal. This larger dataset enhances the scope for studying EEG-based emotional responses.

\subsection{Baseline Methods}
This investigation appraises the effectiveness of our methodology in EEG-based emotion recognition. We benchmark it against a suite of recognized baseline approaches, detailed hereafter:

1) DGCNN (graph-based)\cite{Song2020EEG}: The Dynamical Graph Convolutional Neural Network (DGCNN)  dynamically discerns inter-channel EEG relationships via a trainable adjacency matrix, refined throughout the neural network’s learning process. This adaptability markedly enhances the extraction of discriminative spatial features, bolstering emotion recognition precision.

2) RGNN (graph-based)\cite{RGNN}: The Regularized Graph Neural Network (RGNN)  leverages neuroscientific insights into brain topology to model local and global EEG channel interactions. By embedding sparsity-inducing regularization within its graph convolutions, RGNN prunes extraneous connections, thereby amplifying emotionally salient features and ensuring robust classification across diverse stimuli.

3) PGCN (graph-based)\cite{Jin2015PGCN}: The Pyramidal Graph Convolutional Network (PGCN)  constructs a triadic hierarchy—encompassing local electrode clusters, mesoscopic regions (e.g., seven lobes), and global cortex—using sparse adjacency matrices. This hierarchical synthesis mitigates over-smoothing, yielding a precise and interpretable emotional activity map.

4) TSception (CNN-based)\cite{ding2022tsception}: TSception, a multi-scale convolutional architecture, integrates dynamic temporal, asymmetric spatial, and fusion layers. By concurrently extracting temporal dynamics and spatial asymmetries, it excels in discerning rapid emotional fluctuations across EEG channels.

5) LSTM (temporal-learning)\cite{Soleymani2016}: Long Short-Term Memory (LSTM) networks , equipped with dual memory cells and gating mechanisms, process 4 Hz EEG sequences to capture long-term temporal dependencies. Such capability proves invaluable for tracking gradual emotional transitions, e.g., neutral to positive valence.

6) TCN (temporal-learning)\cite{Zhang2022}: The Temporal Convolutional Network (TCN)  employs adjustable dilated convolutions, augmented by visual-to-EEG distillation, to encapsulate extended temporal patterns, outperforming LSTM in multimodal regression tasks.

7) BiDANN (adversarial-based)\cite{BiDANN}: The Bi-Hemisphere Domain Adversarial Neural Network (BiDANN)  deploys dual-hemisphere LSTM extractors feeding three discriminators, interlinked via Gradient Reversal Layers. This adversarial domain alignment, preserving hemispheric distinctions, ensures robust cross-subject generalization.

8) DMATN (adversarial-based)\cite{wang2021deep}: The Deep Multi-Source Adaptation Transfer Network (DMATN)  synthesizes multi-source EEG through attention-weighted fusion and an adversarial classifier. By harmonizing diverse inputs, it achieves consistent cross-subject performance.

9) EmT (Graph-Transformer-Based)\cite{10960695}: The Emotion Transformer (EmT), a leading method in graph-transformer-based EEG emotion recognition, leverages a graph-transformer architecture to model spatiotemporal dynamics. By converting signals into temporal graphs, its residual multi-view pyramid GCN (RMPG) captures diverse spatial patterns of emotional cognition, while the temporal contextual transformer (TCT) excels at learning long-term dependencies, achieving superior cross-subject generalization in classification and regression tasks.

\begin{table*}[t]
\centering
\small
\setlength{\tabcolsep}{4pt}
\caption{The Accuracies and F1 Scores (mean ± std) on the SEED, THU-EP, and FACED Datasets. The Best Results Are Highlighted in Bold and the Next Best Are Marked Using Underlines.}
\vspace{0.5\baselineskip}
\label{tab:baseline_results}
\begin{tabularx}{\textwidth}{>{\centering\arraybackslash}l*{6}{>{\centering\arraybackslash}X}}
\toprule
\textbf{METHOD} & \multicolumn{2}{c}{\textbf{SEED}} & \multicolumn{2}{c}{\textbf{THU-EP}} & \multicolumn{2}{c}{\textbf{FACED}} \\
\cmidrule(lr){2-3} \cmidrule(lr){4-5} \cmidrule(lr){6-7}
& \textbf{ACC (\%)} & \textbf{F1 (\%)} & \textbf{ACC (\%)} & \textbf{F1 (\%)} & \textbf{ACC (\%)} & \textbf{F1 (\%)} \\
\midrule
KNN        & 49.26 $\pm$ 14.89 & 48.89 $\pm$ 24.91 & 23.45 $\pm$ 4.82 & 30.93 $\pm$ 4.85 & 22.71 $\pm$ 5.34 & 31.89 $\pm$ 4.37 \\
SVM        & 51.68 $\pm$ 17.86 & 50.31 $\pm$ 28.76 & 24.72 $\pm$ 5.91 & 33.28 $\pm$ 8.70 & 25.58 $\pm$ 7.43 & 32.32 $\pm$ 9.23 \\
DGCNN      & 72.48 $\pm$ 14.89 & 61.95 $\pm$ 13.14 & 56.71 $\pm$ 3.37 & 64.74 $\pm$ 5.22 & 56.26 $\pm$ 4.55 & 69.78 $\pm$ 4.36 \\
RGNN       & 79.07 $\pm$ 14.82 & 80.26 $\pm$ 13.34 & 57.23 $\pm$ 3.07 & 69.51 $\pm$ 5.49 & 58.71 $\pm$ 5.05 & 72.27 $\pm$ 72.16 \\
DMATN      & 77.29 $\pm$ 15.46 & 77.16 $\pm$ 16.32 & 60.34 $\pm$ 5.41 & 69.22 $\pm$ 6.38 & 61.41 $\pm$ 4.91 & 68.31 $\pm$ 6.87 \\
TSception  & 66.22 $\pm$ 18.11 & 62.18 $\pm$ 28.34 & 59.18 $\pm$ 5.93 & 70.72 $\pm$ 6.01 & 61.92 $\pm$ 8.81 & 70.26 $\pm$ 23.76 \\
TCN        & 76.54 $\pm$ 14.08 & 73.77 $\pm$ 21.96 & 57.79 $\pm$ 3.13 & 67.71 $\pm$ 3.12 & 55.26 $\pm$ 3.56 & 67.31 $\pm$ 3.53 \\
PGCN       & 75.87 $\pm$ 18.36 & 74.08 $\pm$ 19.01 & 56.92 $\pm$ 4.31 & 65.14 $\pm$ 9.06 & 55.77 $\pm$ 7.81 & 66.54 $\pm$ 8.52 \\
LSTM       & 73.39 $\pm$ 15.88 & 67.05 $\pm$ 27.47 & 55.83 $\pm$ 3.52 & 62.69 $\pm$ 6.22 & 56.84 $\pm$ 6.31 & 70.07 $\pm$ 6.44 \\
BiDANN     & 79.39 $\pm$ 16.45 & 77.65 $\pm$ 15.92 & \underline{61.44 $\pm$ 5.51} & 69.71 $\pm$ 6.87 & \textbf{63.36 $\pm$ 7.01} & 73.82 $\pm$ 6.36 \\
EmT$^*$        & \underline{80.20 $\pm$ 11.50} & \underline{82.10 $\pm$ 9.30} & 59.50 $\pm$ 4.70 & \underline{72.40 $\pm$ 4.40} & 60.80 $\pm$ 6.50 & \underline{74.00 $\pm$ 5.80} \\
\midrule
\textbf{MSGM (Proposed)} & \textbf{83.43 $\pm$ 11.42} & \textbf{85.03 $\pm$ 9.09} & \textbf{62.39 $\pm$ 3.13} & \textbf{73.28 $\pm$ 4.39} & \underline{63.17 $\pm$ 3.62} & \textbf{76.01 $\pm$ 3.74} \\
\bottomrule
\end{tabularx}
\vspace*{0.1\baselineskip} 
\par 
\noindent 
\raggedright 
\textit{* Indicates the model’s results are sourced from the cited paper. Models without * reflect local experimental results.}
\end{table*}

\subsection{Experimental Protocol}
In this study, we adopt a training strategy consistent with the approach in EmT to implement a subject-independent evaluation framework, ensuring effective generalization to unseen individuals across tailored cross-validation strategies for the SEED, THU-EP, and FACED datasets. For the SEED dataset, which includes data from fifteen experimental subjects, we employ a leave-one-subject-out (LOSO) cross-validation approach, where in each of the fifteen iterations, data from one subject are set aside as the test set, and the remaining fourteen subjects' data are pooled and split randomly into training and validation sets at an 8:2 ratio - 80\% for training and 20\% for validation. As previously noted, we sliced the SEED dataset into time windows of varying lengths to serve as input, resulting in data segments of different sizes. To accommodate this variability, we created multiple dataloaders to feed the network, ensuring consistent labeling across all segments. For the THU-EP and FACED datasets, we use a leave-n-subject-out cross-validation strategy, with \( n \) set to 8 for THU-EP and 12 for FACED; in each fold, data from \( n \) subjects form the test set, while the remaining subjects' data are divided so that 90\% go to training and 10\% to validation. Across all three datasets, we classify emotions binarily into positive and negative categories, and for THU-EP and FACED, this involves converting valence scores into high and low categories using a 3.0 threshold. The model is trained on the training set, using the validation set to tune hyperparameters and avoid overfitting, and its performance is assessed on the test set; this process repeats for each fold, with final performance metrics averaged across all iterations.

\subsection{Parameter Settings}

\begin{table}[!t]
\caption{Training Hyperparameters of the MSGM Model}
\centering
\footnotesize
\vspace{0.5\baselineskip}
\begin{tabular}{@{} l c c @{}}
\toprule
\bfseries Hyperparameters & \bfseries Value \\ 
\midrule
Loss function & Cross-entropy \\[0.4ex]
Optimizer & AdamW \\[0.4ex]
Initial learning rate & \( 3 \times 10^{-4} \) \\[0.4ex]
Label smoothing & 0.1 \\[0.4ex]
Dropout rate & 0.25 \\[0.4ex]
Batch size & 32 \\[0.4ex]
Training epochs (SEED) & 20 \\[0.4ex]
Training epochs (THU-EP, FACED) & 30 \\[0.4ex]
Early stopping patience & 5 \\[0.4ex]
\bottomrule
\end{tabular}
\label{tab:train_hyperparameters}
\end{table}

The training configuration of the proposed MSGM model is detailed in Table~\ref{tab:train_hyperparameters}. The model employs cross-entropy loss for optimization, guided by the AdamW optimizer with an initial learning rate of \( 3 \times 10^{-4} \). To address overfitting, label smoothing (0.1) and dropout (0.25) are applied. A batch size of 32 is used across all datasets, with training epochs set to 20 for SEED and 30 for THU-EP and FACED, incorporating an early stopping mechanism with a patience of 5. The model with the highest validation accuracy is selected for testing.

\begin{table}[!t]
\caption{Architectural Hyperparameters of the MSGM Model}
\vspace{0.5\baselineskip}
\centering
\footnotesize
\begin{tabular}{@{} l c c @{}}
\toprule
\bfseries Hyperparameters & \bfseries Value \\ 
\midrule
GCN layers & [1, 2] \\[0.4ex]
Embedding dimension (\( h \)) & 32 \\[0.4ex]
MSST-Mamba layers & 1 \\[0.4ex]
\( dt_{\text{rank}} \) & \( \lceil h / 16 \rceil \) \\[0.4ex]
\( d_{\text{state}} \) & 16 \\[0.4ex]
\bottomrule
\end{tabular}
\label{tab:arch_hyperparameters}
\end{table}

The MSST-Mamba architecture, summarized in Table~\ref{tab:arch_hyperparameters}, leverages two Chebyshev graph encoders with layers [1, 2] to enhance graph processing and handle complex relationships effectively. It employs an embedding dimension \( h = 32 \) and a convolutional kernel size of 4 to capture local temporal patterns efficiently. A single MSST-Mamba layer is adopted for feature extraction and spatiotemporal processing across datasets, achieving high accuracy while maintaining computational efficiency. The selective state-space model (SSM) operates with a dynamically computed \( dt_{\text{rank}} = \lceil h / 16 \rceil \) and a state dimension \( d_{\text{state}} = 16 \), optimizing spatiotemporal modeling.

Hardware configurations are presented in Table~\ref{tab:hardware_specs}. Training and testing leverage an NVIDIA GeForce RTX 3070Ti (8 GB GDDR6), enabling rapid optimization of the model’s parameters. For real-world deployment, the NVIDIA Jetson Xavier NX, featuring a 6-core Carmel ARM v8.2 CPU and a Volta GPU with 48 Tensor Cores (up to 21 TOPS, INT8), offers low-power (10–20 W) and high-efficiency inference, supported by 8 GB LPDDR4x memory and 51.2 GB/s bandwidth, ideal for edge computing applications.

\begin{table}[!t]
\caption{Hardware Specifications for Training and Deployment}
\vspace{0.5\baselineskip}
\centering
\footnotesize
\begin{tabular}{@{} l c c @{}}
\toprule
\bfseries Property & \bfseries GeForce RTX 3070Ti & \bfseries Jetson Xavier NX \\
\midrule
GPU & NVIDIA GeForce RTX 3070Ti & NVIDIA Volta \\ 
\addlinespace[0.5ex]
CPU & Core i7-8700K & Carmel Arm v8.2 \\ 
\addlinespace[0.5ex]
RAM & 64 GB & 8 GB \\ 
\addlinespace[0.5ex]
Power usage & 240 W & 10 W / 15 W / 20 W \\ 
\addlinespace[0.5ex]
Purpose & Training and testing & Real-world deployment \\ 
\bottomrule
\end{tabular}
\label{tab:hardware_specs}
\end{table}

\section{Numerical Results}
\subsection{Emotion Recognition Performance}
The experimental results are presented in Table~\ref{tab:baseline_results}, which evaluates the performance of various methods for generalized emotion classification across three datasets—SEED, THU-EP, and FACED—using accuracy (ACC \%) and F1 score (F1 \%) as metrics. On the SEED dataset, our proposed method achieves an outstanding accuracy of 83.43\% and an F1 score of 85.03\%, outperforming all other approaches. The next best performers are EmT with an accuracy of 80.20\% and an F1 score of 82.10\% (second-highest accuracy and F1 score), while traditional methods like KNN and SVM trail far behind with accuracies of 49.26\% and 51.68\%, and F1 scores of 48.89\% and 50.31\%, respectively. For the THU-EP dataset, our method continues to lead with the highest accuracy of 62.39\% and F1 score of 73.28\%, followed closely by BiDANN at 61.44\% accuracy and EmT at 72.40\% F1 score. On the FACED dataset, BiDANN achieves a slightly higher accuracy of 63.36\% compared to our method's 63.17\%. However, our method outperforms BiDANN in terms of F1 score, achieving 76.01\% against BiDANN's 73.82\%, making our model the top performer in F1 score.

Across all three datasets, our proposed method consistently delivers superior performance, particularly excelling in accuracy, thereby affirming its robustness and effectiveness in emotion classification tasks. Advanced architectures such as EmT and BiDANN also demonstrate strong capabilities; for instance, EmT achieves the second-best accuracy on SEED and competitive results on THU-EP and FACED, while BiDANN stands out with the second-highest accuracy on THU-EP. Methods leveraging temporal dynamics, such as TSception (66.22\% ACC on SEED) and TCN (76.54\% ACC on SEED), generally outperform those relying solely on spatial features, highlighting the critical role of temporal information in EEG-based emotion recognition. The substantial performance gap between these advanced methods and traditional approaches like KNN and SVM underscores the limitations of simpler models in this complex domain.

\begin{table}[!t]
\caption{Generalized Emotion Classification Results of Ablation Studies on the SEED and THU-EP Datasets (\%)}
\vspace{0.5\baselineskip}
\centering
\footnotesize
\begin{tabular}{@{} l c c c c @{}}
\toprule
\textbf{Method} & \multicolumn{2}{c}{\textbf{SEED}} & \multicolumn{2}{c}{\textbf{THU-EP}} \\
\cmidrule(lr){2-3} \cmidrule(lr){4-5}
 & \textbf{ACC (\%)} & \textbf{F1 (\%)} & \textbf{ACC (\%)} & \textbf{F1 (\%)} \\
\midrule
w/o Temporal Multi-Scale & 80.04 & 80.14 & 59.55 & 72.13 \\ 
\addlinespace[0.5ex]
w/o Spatial Multi-Scale & 82.37 & 82.57 & 61.82 & 72.02 \\ 
\addlinespace[0.5ex]
w Single GCN & 81.92 & 81.50 & 62.01 & 70.68 \\ 
\addlinespace[0.5ex]
w/o Spatiotemporal Fusion & 79.75 & 80.23 & 59.17 & 66.10 \\ 
\addlinespace[0.5ex]
w/o MSST-Mamba & 79.53 & 77.93 & 57.92 & 65.04\\ 
\addlinespace[0.5ex]
MSGM (Proposed) & \textbf{83.43} & \textbf{85.03} & \textbf{62.39} & \textbf{73.28} \\ 
\bottomrule
\end{tabular}
\label{tab:ablation_results}
\end{table}

\subsection{Ablation Study on Component Modules}
To evaluate the contributions of the temporal multi-scale feature extraction, spatial multi-scale prior information initialization, spatiotemporal feature adaptive fusion, and MSST-Mamba and classifier modules, we conducted an ablation analysis by systematically removing each component individually and assessing its impact on classification performance. This included omitting the temporal multi-scale feature extraction (w/o Temporal Multi-Scale), spatial multi-scale prior information initialization (w/o Spatial Multi-Scale), spatiotemporal feature adaptive fusion (w/o Spatiotemporal Fusion), and MSST-Mamba (w/o MSST-Mamba), as well as replacing multi-depth GCNs with a single layer (w Single GCN), to measure each component’s effect. The results are detailed in Table~\ref{tab:ablation_results}.

The removal of the MSST-Mamba and classifier module results in the most significant performance decline, with accuracy decreasing by 3.90\% on the SEED dataset and 4.47\% on the THU-EP dataset, alongside F1 score drops of 7.10\% and 8.24\%, respectively. This underscores its critical role in processing and integrating multi-scale spatiotemporal features effectively. Excluding the spatiotemporal feature adaptive fusion module also leads to substantial reductions, with accuracy dropping by 3.68\% on SEED and 3.22\% on THU-EP, highlighting its importance in unifying temporal and spatial information. 

The absence of the temporal multi-scale feature extraction module decreases accuracy by 3.39\% on SEED and 2.84\% on THU-EP, indicating its value in capturing diverse temporal dynamics. Removing the spatial multi-scale prior information initialization module results in smaller but notable declines of 1.06\% on SEED and 0.57\% on THU-EP, suggesting its contribution to initializing robust spatial representations, though its impact is less pronounced than other modules. Additionally, using a single GCN instead of multiple GCN layers reduces accuracy by 1.51\% on SEED and 0.38\% on THU-EP, demonstrating that multi-layer GCNs more effectively capture spatial information.

\subsection{Performance and Sensitive Analysis of Hyperparameters}
\subsubsection{Impact of EEG Feature Types}
\begin{figure*}[t!]
    \centering
    \begin{subfigure}[b]{0.31\textwidth}
        \includegraphics[width=\textwidth]{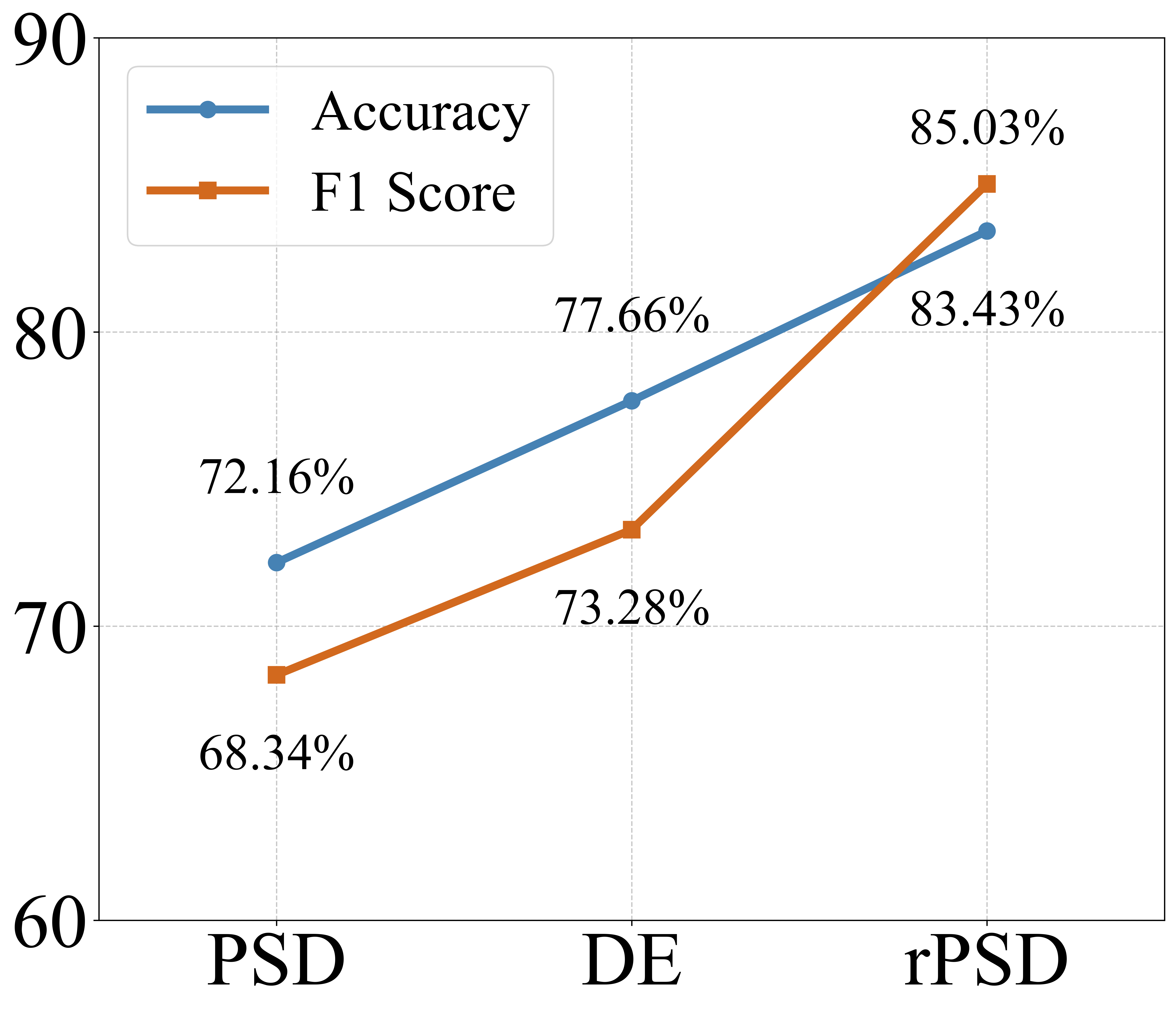}
        \caption{}
        \label{fig:rPSDvsPSDvsDE}
    \end{subfigure}
    \hfill
    \begin{subfigure}[b]{0.31\textwidth}
        \includegraphics[width=\textwidth]{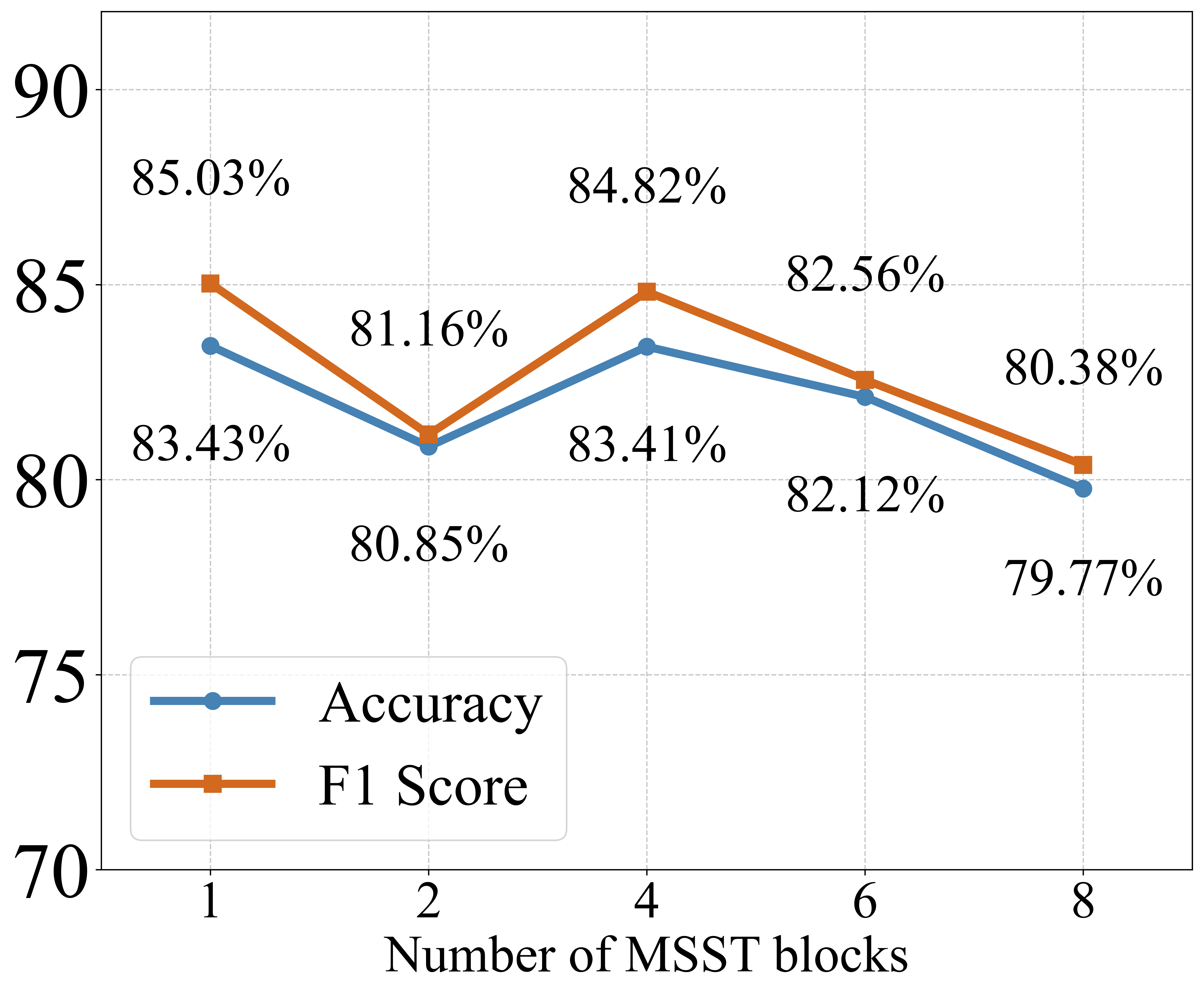}
        \caption{}
        \label{fig:layer_num}
    \end{subfigure}
    \hfill
    \begin{subfigure}[b]{0.31\textwidth}
        \includegraphics[width=\textwidth]{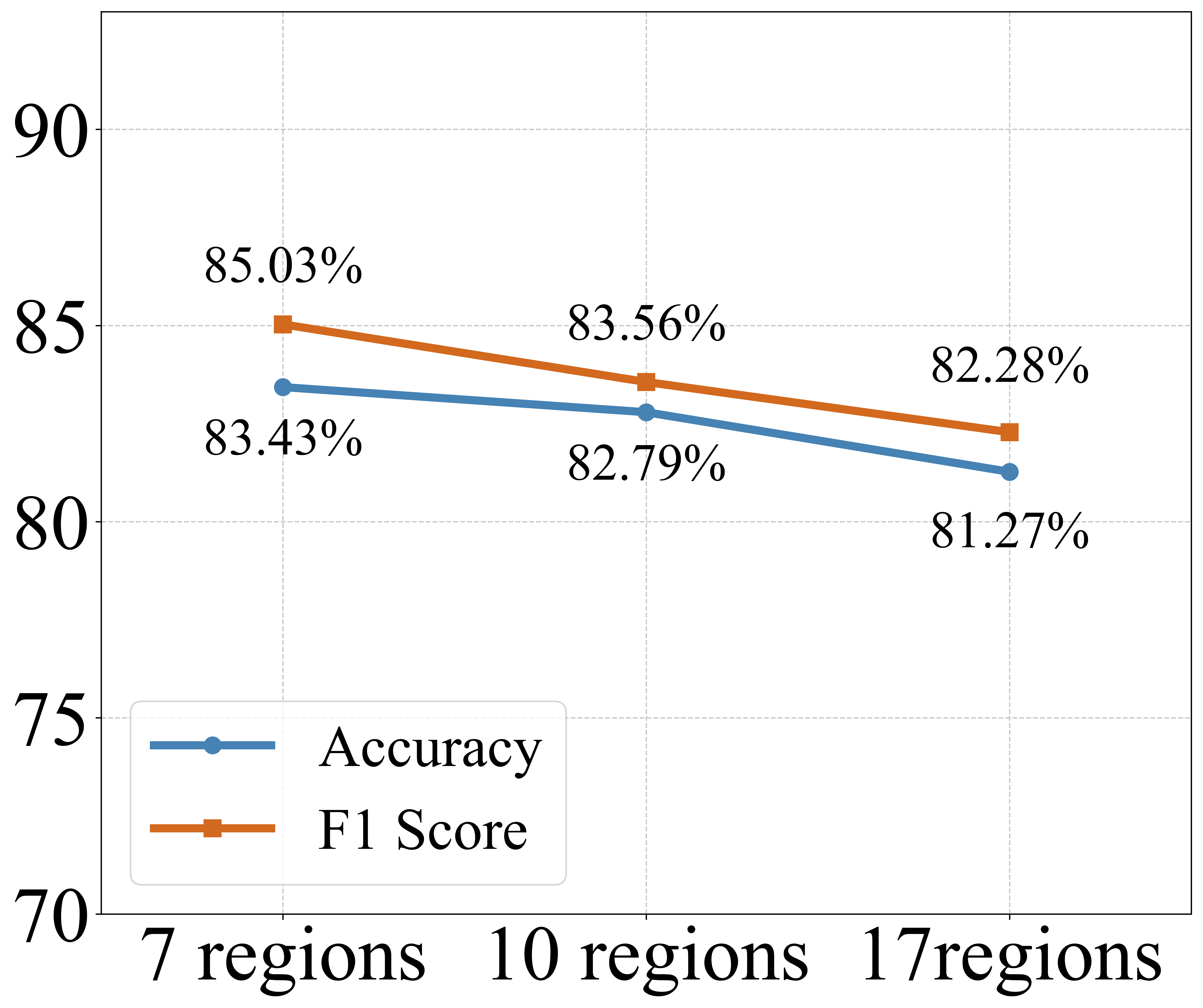}
        \caption{}
        \label{fig:region_num}
    \end{subfigure}
    \caption{(a) Effect of feature types on emotion classification performances using SEED. (b) Effect of the number of MSST blocks on emotion classification using SEED. (c) Effect of different brain region partitioning methods on emotion recognition using SEED.}
    \label{fig:overall1}
\end{figure*}

As illustrated in Figure~\ref{fig:rPSDvsPSDvsDE}, which compares the accuracy and F1 scores of Power Spectral Density (PSD), Differential Entropy (DE), and relative Power Spectral Density (rPSD) features for emotion classification on the SEED dataset. Specifically, rPSD achieved an accuracy of 83.43\% and an F1 score of 85.03\%, surpassing DE by 5.77 percentage points in accuracy and 11.75 percentage points in F1 score. Compared to PSD, rPSD exhibited even greater improvements, with an accuracy increase of 11.27 percentage points and an F1 score increase of 16.69 percentage points. These findings demonstrate that rPSD is a superior feature in our model for EEG-based emotion classification tasks compared to both DE and PSD.

\subsubsection{Influence of the Number of MSST-Mamba Blocks}
The effect of varying the number of MSST-Mamba Blocks on the emotion classification performance is illustrated in Figure~\ref{fig:layer_num}. The analysis considered block counts of 1, 2, 4, 6, and 8, with corresponding impacts on accuracy and F1 score. With a single block, the model achieved an accuracy of 83.43\% and an F1 score of 85.03\%. Increasing to 2 blocks led to a decrease in performance, with accuracy dropping to 80.85\% and F1 score to 81.16\%. A slight recovery was observed with 4 blocks, where accuracy reached 83.41\% and F1 score 84.82\%, nearly matching the single-block performance. Further increasing the block count to 6 resulted in a decline, with accuracy at 82.12\% and F1 score at 82.56\%, and this downward trend persisted with 8 blocks, where accuracy and F1 score further decreased to 79.77\% and 80.38\%, respectively. This pattern suggests that a single block achieves the best performance, with additional blocks leading to fluctuations and an overall decline at higher counts.

\subsubsection{Effect of Prior Information on Brain Region Partitioning}
\begin{figure}[t!]
\centering
\includegraphics[width=\columnwidth]{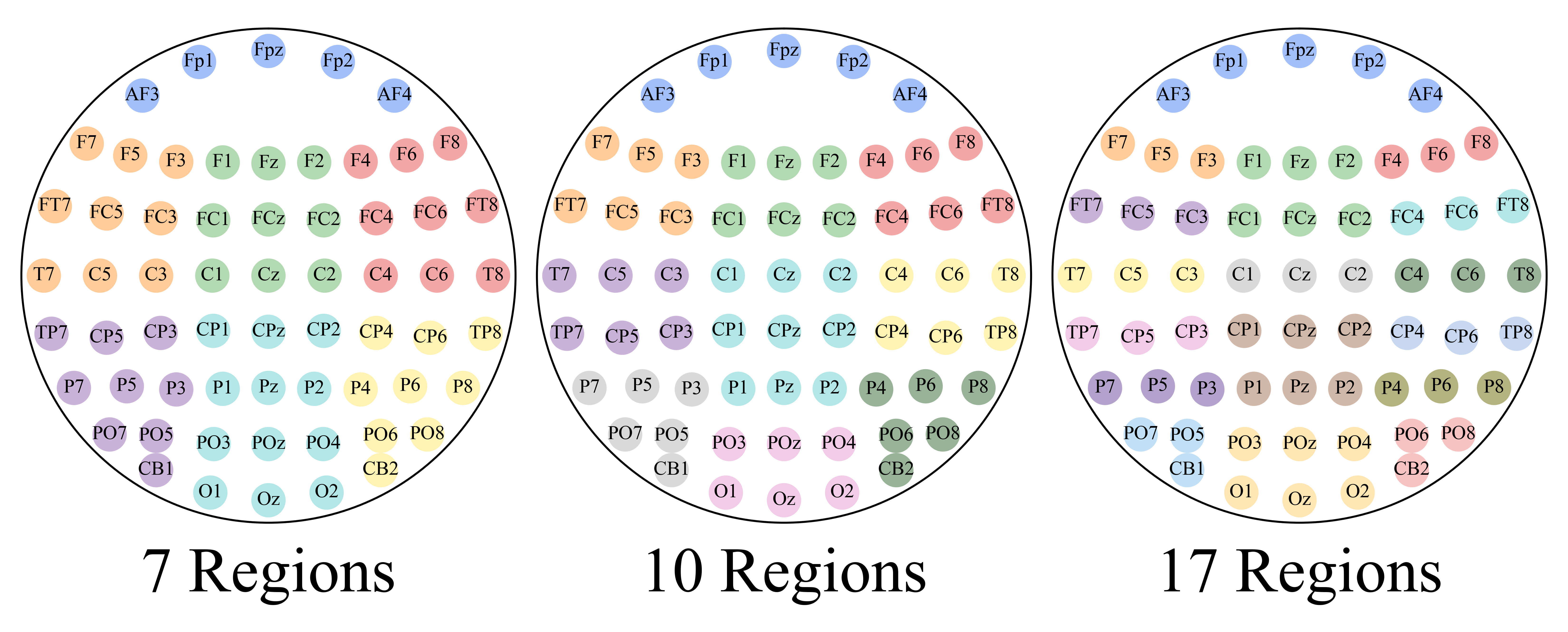}
\caption{Three methods for dividing 62 EEG channels into different regions, comprising 7, 10, and 17 regions respectively.}
\label{fig:region_figure}
\end{figure}

The human brain comprises multiple functional regions, each contributing uniquely to emotional processing\cite{alarcao2017emotions}. The way these regions are partitioned into subgraphs can influence the structure of the EEG data representation and, consequently, the model’s performance\cite{song2021variational}. To explore this, we conducted experiments on the SEED dataset by dividing the 62 EEG channels into 7, 10, and 17 regions, as shown in Figure~\ref{fig:region_figure}. Our results indicate (see Figure~\ref{fig:region_num}) that the 7-region partitioning yields the highest accuracy (83.43\%) and F1 score (85.03\%), followed by the 10-region partitioning with an accuracy of 82.79\% and an F1 score of 83.56\%, while the 17-region partitioning produces the lowest accuracy (81.27\%) and F1 score (82.28\%). These findings suggest that the 7-region scheme may strike an effective balance between capturing essential functional patterns and maintaining a manageable level of complexity for the model. In contrast, the finer 17-region partitioning might overly fragment the data, diluting key inter-regional relationships, while the 10-region approach, despite performing better than 17 regions, may still not align as optimally with the underlying functional organization as the 7-region configuration.

\begin{figure}[htbp]
    \centering
    \begin{subfigure}[b]{\columnwidth}
        \includegraphics[width=\textwidth]{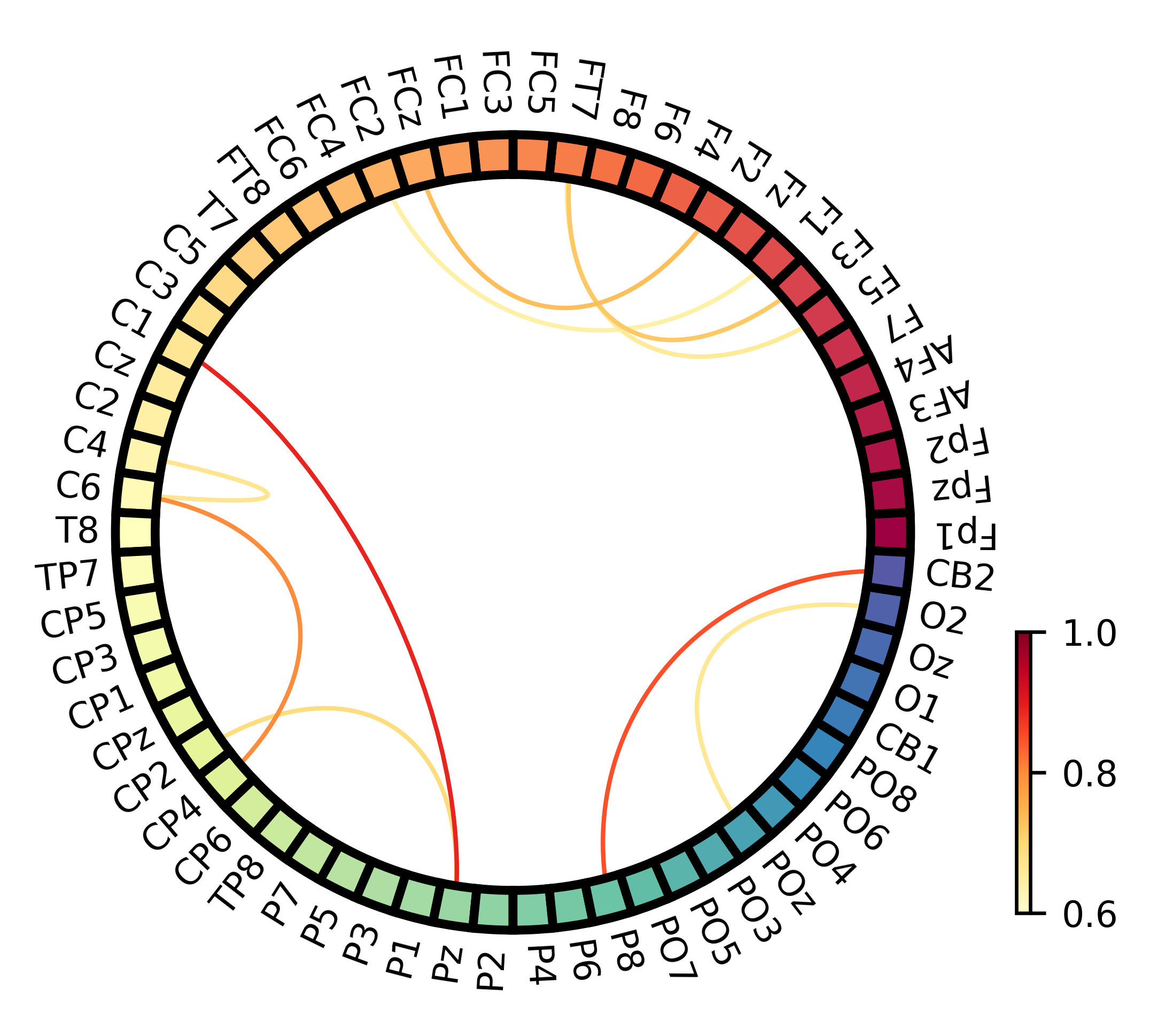}
        \caption{}
        \label{fig:connectivity_map_init}
    \end{subfigure}
    
    \begin{subfigure}[b]{\columnwidth}
        \includegraphics[width=\textwidth]{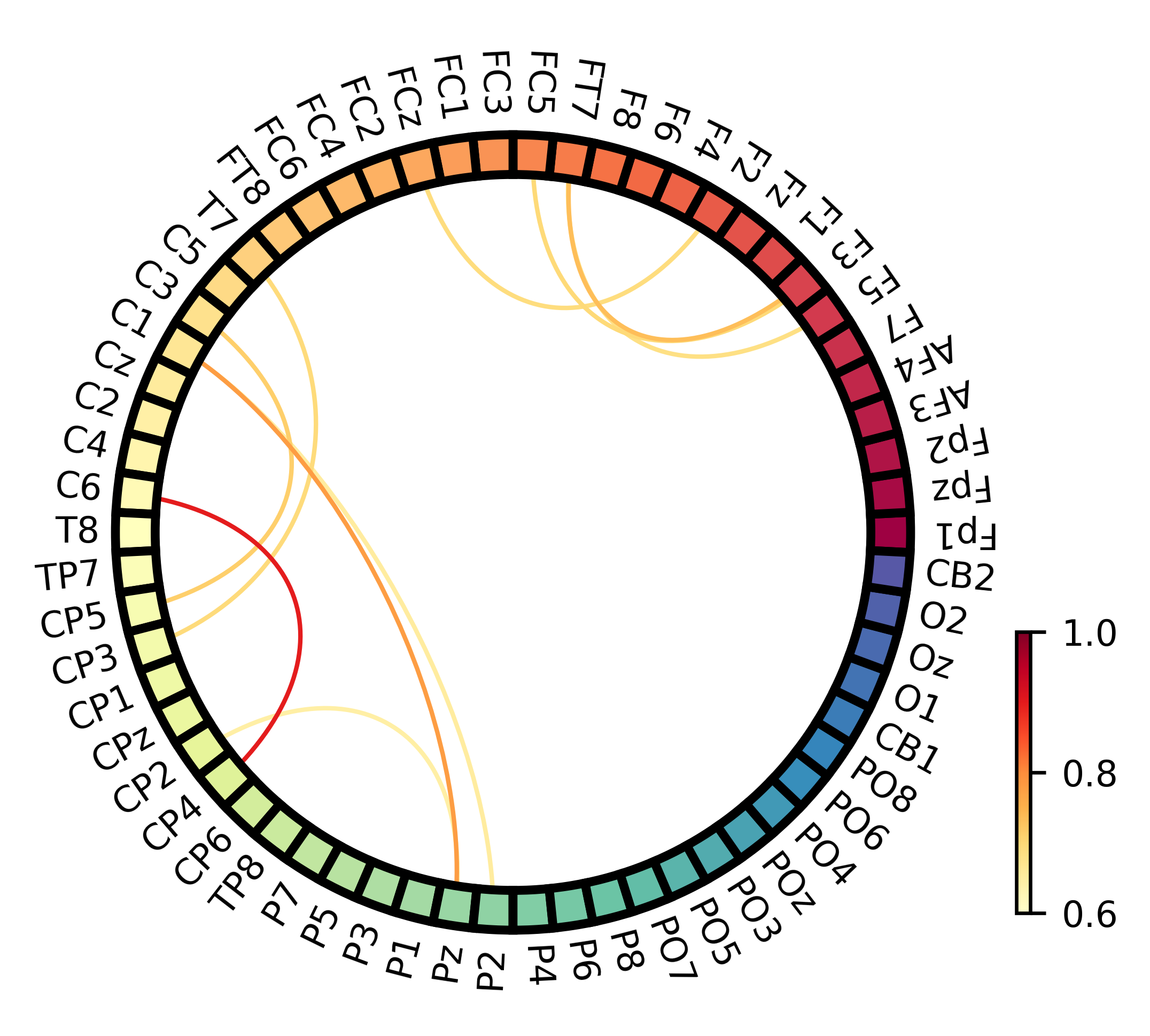}
        \caption{}
        \label{fig:connectivity_map}
    \end{subfigure}
    
    \caption{(a) Connectivity between the electrodes of the initial Local Graph. (b) Connectivity between the electrodes of the trained Local Graph.}
    \label{fig:overall2}
\end{figure}

\begin{figure}[h]
    \centering
    \includegraphics[width=0.3\textwidth]{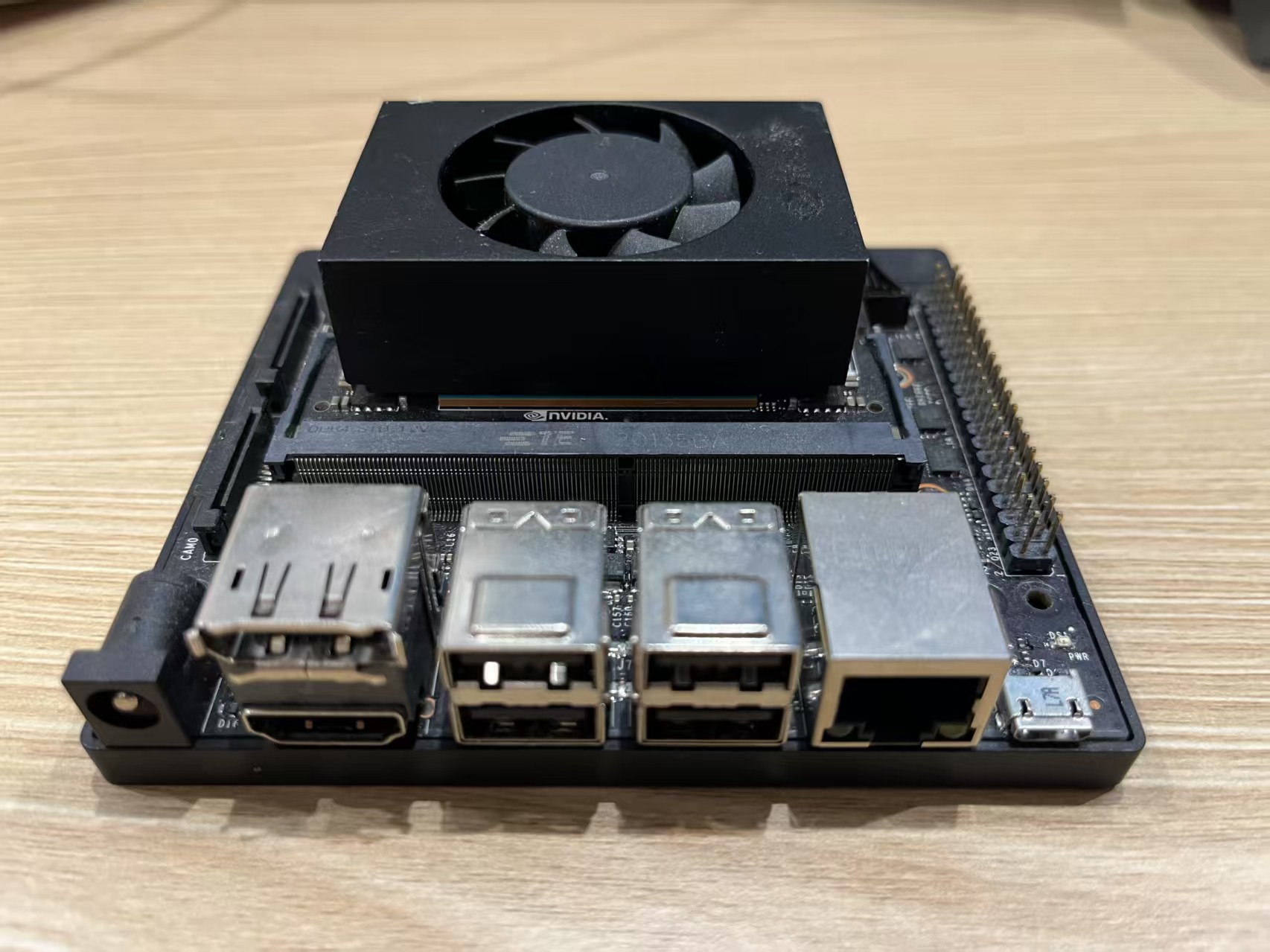}
    \caption{NVIDIA Jetson Xavier NX.}
    \label{fig:NVIDIA_Jetson_Xavier_NX}
\end{figure}

\subsection{Performance on Edge Devices}
The MSGM model, deployed on the NVIDIA Jetson Xavier NX edge computing platform, as shown in Figure~\ref{fig:NVIDIA_Jetson_Xavier_NX}, exhibits efficient performance on the SEED dataset. To enable deployment on this platform, we replaced the Mamba core component in the MSST-Mamba module with Mamba-minimal\footnote{\href{https://github.com/johnma2006/mamba-minimal}{https://github.com/johnma2006/mamba-minimal}}, a lightweight implementation of Mamba, since the PyTorch version on the edge device does not support the official Mamba library. This substitution preserves the model’s input-output functionality but results in lower runtime efficiency compared to the official Mamba implementation. With this configuration, the model utilizes 349,218 parameters and achieves an inference time of 151.0 ms, maintaining millisecond-level inference and demonstrating robust real-time processing capabilities. This efficiency underscores its suitability for edge device applications requiring rapid data handling.

\subsection{Comparison with EmT}
In this section, we compare our MSGM model with EmT, a leading method in graph-transformer-based EEG emotion recognition. Both models adopt a graph-Transformer/Mamba-based architecture to process spatial-temporal patterns in EEG signals. EmT incorporates an 8-layer TCT module, while MSGM employs a single-layer MSST-Mamba module. We evaluate their performance in terms of accuracy, parameter count, and inference time. To ensure a fair comparison, this evaluation was conducted on the GeForce RTX 3070Ti platform (see Table~\ref{tab:hardware_specs}), rather than on edge devices, allowing both MSGM and EmT to run in a consistent environment without the influence of Mamba-minimal, which was used for edge deployment.

MSGM, with its single-layer MSST-Mamba, achieves superior accuracy and F1 scores compared to EmT, despite using only 349,218 parameters—approximately half of EmT’s 703,530. This highlights MSGM’s efficiency, as its linear-complexity MSST-Mamba outperforms the quadratic-complexity TCT module with a simpler structure. The reduced parameter count underscores MSGM’s suitability for resource-constrained settings, such as edge devices.

In terms of inference time, MSGM records 7.9 ms, slightly higher than EmT’s 4.3 ms. This minor gap arises from MSGM’s multi-scale architecture, which limits full parallelization. Nevertheless, both models maintain millisecond-level inference, ensuring negligible impact on real-time applications.

\subsection{Visualization}
Figure~\ref{fig:overall2} presents two diagrams that illustrate the connectivity between different electrodes, derived from the initial and learned perspectives, respectively. We utilized a Local Graph derived from the Global Graph as the representation, which effectively reflects the model’s learning outcomes. These diagrams highlight the evolution of connectivity patterns identified during the training process, based on data from the EEG dataset.

In the initial connectivity map (Figure~\ref{fig:connectivity_map_init}), the strongest connections are observed between electrodes such as C1-Pz, FC2-FPz, and C6-CP4. These connections primarily involve the central and parietal regions, with some involvement of the frontal areas\cite{wan2021frontal}, suggesting a baseline interaction that may reflect general neural communication prior to task-specific learning. The prominence of these connections indicates an initial focus on central-parietal and frontal-central interactions, which are often associated with sensory and motor coordination in early-stage processing.

In contrast, the learned connectivity map (Figure~\ref{fig:connectivity_map}) reveals a more refined set of connections, with the strongest links being C6-CP4 and C1-Pz. These retained and strengthened connections continue to emphasize interactions within the central and parietal regions, which are known to play critical roles in sensory integration and spatial processing. The persistence of these specific connections suggests that the model has prioritized and enhanced these pathways, likely due to their relevance to the task at hand\cite{alqazzaz2019electroencephalogram}. Additionally, the color intensity, ranging from 0.6 to 1.0, highlights the varying strengths of these learned connections, with warmer colors indicating stronger interactions.

The connectivity patterns observed in Figure~\ref{fig:overall2} demonstrate the model's ability to refine and focus on key electrode relationships, transitioning from a broader initial state to a more targeted, task-driven network. This evolution underscores the model's effectiveness in capturing and enhancing critical neural relationships, particularly in the central and parietal regions, tailored to the cognitive demands of the classification task.

\section{Conclusion}
In this paper, we propose the Multi-Scale Spatiotemporal Graph Mamba (MSGM), a novel framework for EEG-based emotion recognition that integrates temporal multi-scale feature extraction, spatial multi-scale prior information initialization, spatiotemporal feature adaptive fusion, and the MSST-Mamba module. By capturing short-term emotional continuity and long-term evolutionary trends through multi-scale temporal analysis, alongside hierarchical spatial connectivity via bimodal graph modeling, MSGM addresses critical gaps in prior methodologies. Extensive experiments on the SEED, THU-EP, and FACED datasets demonstrate its superior performance over baseline methods, validated through rigorous subject-independent evaluation. The model achieves millisecond-level inference speed on edge devices like the NVIDIA Jetson Xavier NX, underscoring its practical applicability in clinical and consumer settings, while its neuroanatomical grounding enhances interpretability of the brain’s distributed emotional dynamics. Nevertheless, challenges remain in achieving optimal cross-subject generalization, constrained by variability in EEG data across individuals. Looking ahead, future developments in EEG-based emotion recognition could focus on integrating multimodal physiological signals—such as ECG or eye-tracking data—to enrich emotional context, optimizing lightweight architectures for broader deployment on wearable devices, and exploring real-time adaptive learning to dynamically adjust to individual neurophysiological profiles, thereby advancing both precision and accessibility in real-world applications.

\normalsize
\bibliography{ref}
\bibliographystyle{IEEEtran}

\end{document}